\documentclass[10pt]{iopart}
\usepackage[comma,square,sort&compress,numbers]{natbib}
\usepackage{graphicx}
\expandafter\let\csname equation*\endcsname\relax
\expandafter\let\csname endequation*\endcsname\relax
\usepackage{amsmath}
\def\Kone{K_\mathrm{\uppercase\expandafter{\romannumeral1}}}
\def\Ktwo{K_\mathrm{\uppercase\expandafter{\romannumeral2}}}

\begin{document}

\title[]{A Planning-and-Exploring Approach to Extreme-Mechanics Force Fields}

\author{Pengjie Shi, Zhiping Xu}

\address{Applied Mechanics Laboratory and Department of Engineering Mechanics, Tsinghua University, Beijing, 100084, China} 
\ead{xuzp@tsinghua.edu.cn}
\vspace{10pt}
\begin{indented}
\item[]\today
\end{indented}

\begin{abstract}
Extreme mechanical processes such as strong lattice distortion and bond breakage during fracture are ubiquitous in nature and engineering, which often lead to catastrophic failure of structures.
However, understanding the nucleation and growth of cracks is challenged by their multiscale characteristics spanning from atomic-level structures at the crack tip to the structural features where the load is applied.
Molecular simulations offer an important tool to resolve the progressive microstructural changes at crack fronts and are widely used to explore processes therein, such as mechanical energy dissipation, crack path selection, and dynamic instabilities (e.g., kinking, branching).
Empirical force fields developed based on local descriptors based on atomic positions and the bond orders do not yield satisfying predictions of fracture, even for the nonlinear, anisotropic stress-strain relations and the energy densities of edges.
High-fidelity force fields thus should include the tensorial nature of strain and the energetics of rare events during fracture, which, unfortunately, have not been taken into account in both the state-of-the-art empirical and machine-learning force fields.
Based on data generated by first-principles calculations, we develop a neural network-based force field for fracture, NN-F$^3$, by combining pre-sampling of the space of strain states and active-learning techniques to explore the transition states at critical bonding distances.
The capability of NN-F$^3$ is demonstrated by studying the rupture of h-BN and twisted bilayer graphene as model problems.
The simulation results confirm recent experimental findings and highlight the necessity to include the knowledge of electronic structures from first-principles calculations in predicting extreme mechanical processes.
\end{abstract}

%
\noindent{\it Keywords}: Machine-learning force fields, Fracture, h-BN, 2D materials, Molecular dynamics

\submitto{\JPCM}
%
\maketitle
%
\ioptwocol

\section{Introduction}
Mechanical properties of materials in extreme strain states are sensitive to flaws\,\cite{ritchie_2011}.
Defects and voids create stress concentration and reduce the strength and strain to failure\,\cite{cao2020elastic, han2020large}.
Cracks with atomically sharp fronts are more detrimental defects, which result in stress singularity in the framework of continuum mechanics.
In the Griffith theory, the stability of cracks is governed by an energy balance between the elastic strain energy stored in the structure and the energy penalty of newly created surfaces.
The $r^{-1/2}$ divergence of the stress and strain fields predicted by the theory does not apply in realistic materials where atoms are arranged in lattices and the bonds between atoms have finite strengths.
Thermodynamics of fracture can be empirically formulated in terms of continuum mechanics.
However, kinetics at the crack tip has to be discussed by considering the lattice discreteness and the transition states of bond breakage and (re)formation.
Molecular simulations offer a powerful tool to address these issues, which provides consistent predictions with the continuum theory at the structural level\,\cite{buehler_2010} and, at the same time, unveils atomic-level kinetics such as lattice trapping processes that are validated by recent \emph{in situ} electron microscopy observations conducted\,\cite{zhigong_2021,ly_2017,huang_sciadv_2020,huang_prl_2020}.

\begin{figure*}[t]%
\centering
\includegraphics[width=\textwidth]{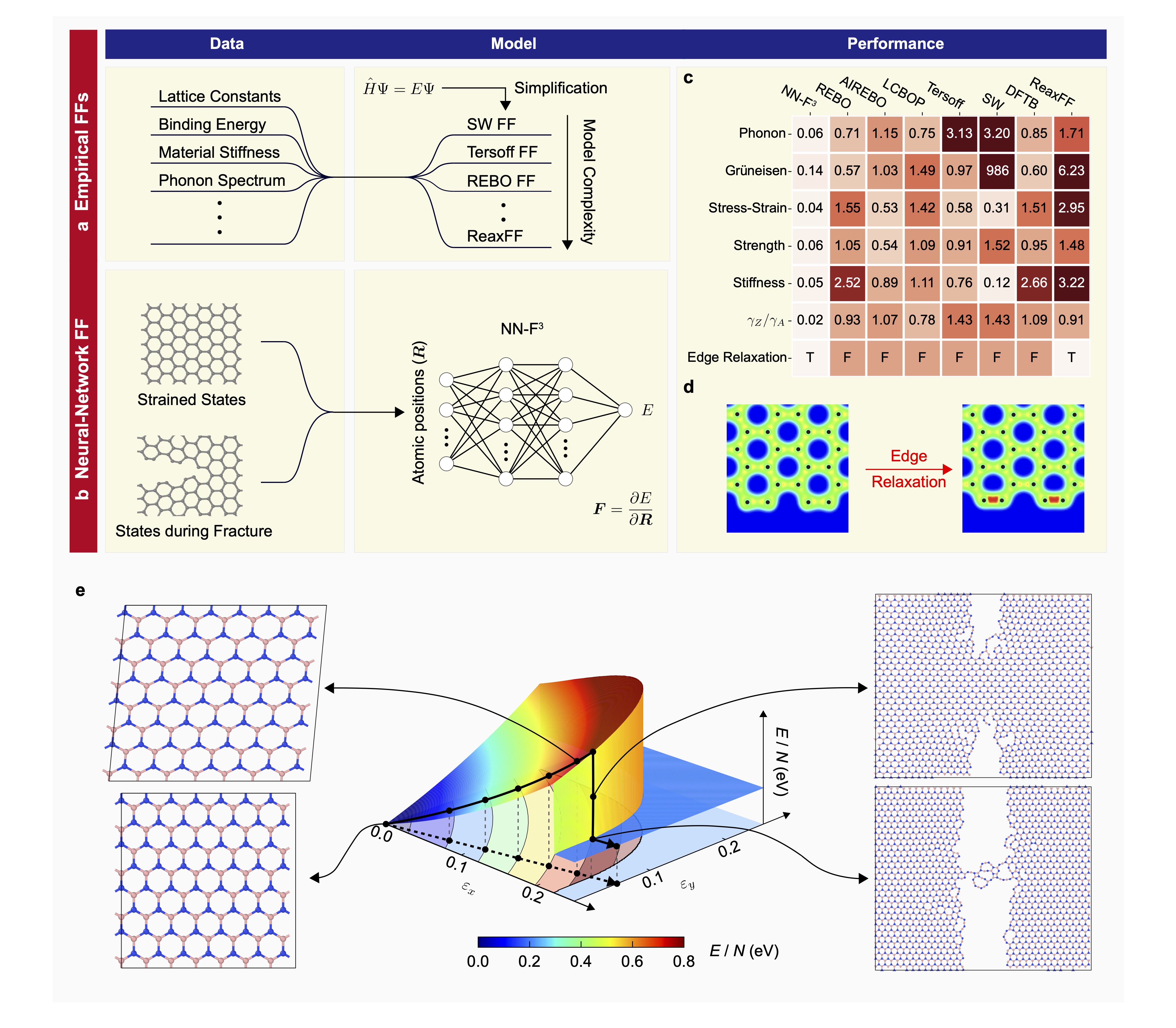}
\caption{
    Construction of empirical\,\cite{hossain_2018,brenner2002second,stuart2000reactive,los2003intrinsic,lindsay2010optimized,jensen2015simulation,hourahine2020dftb} and neural network force fields (FFs).
    \textbf{a}, \textbf{b}, Procedures of parameterization, which include `Data', `Model', and `Performance'.
    For empirical FFs (\textbf{a}), data obtained from first-principles calculations such as lattice constants are used to parameterize the models.
    For NN-F$^3$ (\textbf{b}), strained structures and crack tips are used to train the FF.
    \textbf{c}, Performance of the FFs quantified by the absolute errors (AEs) of predictions, where dark (light) colors measure the AEs. 
    \textbf{d}, Charge density distribution at graphene edges with unrelaxed and relaxed structures.
    \textbf{e}, Potential energy surface (PES) of h-BN with strained and fractured structures. $E$ and $N$ are the total energy and total number of atoms, respectively.
}
\label{Fig_1}
\end{figure*}

First-principles calculations at the Hartree-Fock (HF) or density functional theory (DFT) levels can model ground-state properties of materials based on low-order approximations for the many-body problems of electrons.
The properties of crystals and their defects can be predicted with high accuracy.
However, the heavy computational cost and poor scaling with increasing the number of electrons make it challenging to meet the size and time requirements in simulating extreme mechanical processes such as the dynamics of fracture\,\cite{kermode2008low, buehler_2010}. 
To overcome this problem, empirical force fields (FFs) were developed as an alternative description of the interatomic interaction, which employ simplified representations of the underlying electronic structures of materials in terms of interatomic distances and additional features such as their bond orders\,\cite{hossain_2018,brenner2002second,stuart2000reactive,los2003intrinsic,lindsay2010optimized,jensen2015simulation}.
However, parameterization of empirical FFs is usually based on the equilibrium properties derived from first-principles calculations and/or experimental data and extrapolated to extreme mechanical processes\,\cite{tersoff1988new} (Fig.\,\ref{Fig_1}a).
The large-amplitude and non-uniform strain states at the crack tip as well as the non-equilibrium processes during crack development cannot be well captured\,\cite{hossain_2018}, resulting in lower fidelity compared to first-principles calculations (Fig.\,\ref{Fig_1}c,d)\,\cite{atrash2011evaluation}.
Notable examples include that the nonlinear, anisotropic stress-strain relations under large strain cannot be reproduced, and the edge energy density of 2D crystals cannot be reasonably predicted~\cite{liu2010graphene,feng_2022,qu_2022}.
Efforts to improve the capability of empirical FFs led to the development of more complex representations by adding physics such as charge equilibration in, e.g., the reactive FFs (ReaxFFs)\,\cite{senftle2016reaxff}.
However, the implementations are usually limited by the transferability of parameters that are usually fitted to specific systems\,\cite{van2015development, jensen2015simulation} (Supplementary Fig.\,2a,b).

Recently, the dilemma between accuracy and efficiency in molecular simulations has been tackled by developing machine-learning FFs (MLFFs), for example, using artificial neural networks (NNs)\,\cite{friederich_2021}.
Instead of using interatomic potential functions with explicit mathematical forms as in empirical FFs (e.g., Lennard-Jones, Morse, Stillinger-Weber\,\cite{hossain_2018}, Tersoff\,\cite{lindsay2010optimized}, Brenner\,\cite{lindsay2010optimized}), models such as NNs offer more flexible and efficient ways to map the atomic-level structures of materials to their potential energies and forces\,\cite{deepmd}. 
In practice, MLFFs can be trained using massive data computed from quantum chemistry (e.g., HF, coupled cluster singles and doubles or CCSD), DFT or quantum Monte Carlo (QMC) calculations\,\cite{anderson2022e} (Fig.\,\ref{Fig_1}b).
This approach offers insights into the material behaviors and physicochemical processes\,\cite{zeng2020complex} and opens up dimensions of research in the field of material\,\cite{yin2021atomistic,li2022origin}, environmental\,\cite{galib2021reactive} and chemical\,\cite{vandermause2022active} sciences.

To reach the first-principles level accuracy in modeling mechanical behaviors of materials under extreme strain conditions such as that during fracture, specific considerations should be made in preparing training data for developing MLFFs.
The tensorial nature of stress or strain states near the crack tips should be included with purpose.
Even the basal-plane stress or strain in 2D materials spans over a 3D space ($\sigma_{\alpha\beta}$ or $\varepsilon_{\alpha\beta}$, $\alpha, \beta = x, y)$.
On the other hand, rare events such as bond breakage and (re)formation during cracking are also of crucial importance, especially for capturing the lattice excitation and relaxation processes (Fig.\,\ref{Fig_1}e).
Unfortunately, these features have not yet been implemented in existing MLFFs.
As a result, reported stress-strain relations and predicted fracture patterns show significant deviation from the reference calculations\,\cite{mortazavi_2021,zhang2022atomistic}.
Specifically, the relative errors in strength could reach 20\% at some direction of fracture\,\cite{mortazavi_2021}.

In this article, we address these issues by proposing a planning-and-exploring approach, where pre-sampling of the strain states and active learning are integrated to explore the extreme strain states of materials during fracture. 
A neural network-based force field for fracture (NN-F$^3$) is developed to capture both equilibrium and non-equilibrium features of crack-containing 2D crystals in high fidelity.
The capability of NN-F$^3$ is demonstrated by two representative examples showing the fracture patterns and toughening mechanisms of 2D monolayers and bilayers.

\section{Methods} 
Fracture is usually nucleated at stress concentrators in materials, spanning over multiple length and time scales\,\cite{buehler_2010}.
The growth of existing cracks is controlled by the extreme strain states at the crack tip, where bond breakage occurs.
Edges or surfaces are then cleaved and relaxed\,\cite{kermode2008low}.
The degree of strain and stress concentration at the crack tip depends largely on the loading conditions and sample geometry for specific materials.
Bond breakage and (re)formation are atomistic events that are sensitive to the local environment and can only be accurately captured by models that take into account the ground-state electronic structures by quantum simulations\,\cite{buehler2007threshold}.
Consequently, to simulate the fracture of materials using MLFFs, the stress-strain relations in the tensorial form and crack-tip kinetics have to be included in the training set (Fig.\,\ref{Fig_1}e).
We choose h-BN and graphene as two representative materials for their partly ionic and covalent nature of bonding, respectively.
Their atomic-level structures in the honeycomb lattice are simple, but their mechanical behaviors are rich\,\cite{zhang_2014,feng_2022,qu_2022,zhigong_2021}.
We pre-sample the 3D space of basal-plane strain states and construct an active-learning framework to explore the rare events during cracking using the Deep Potential Smooth Edition (DeepPot-SE)\,\cite{NEURIPS2018_e2ad76f2} model.
The methodology is implemented in DeePMD-kit\,\cite{deepmd} to develop the NN-F$^3$.

\begin{figure*}[t]
\centering
\includegraphics[width=\textwidth]{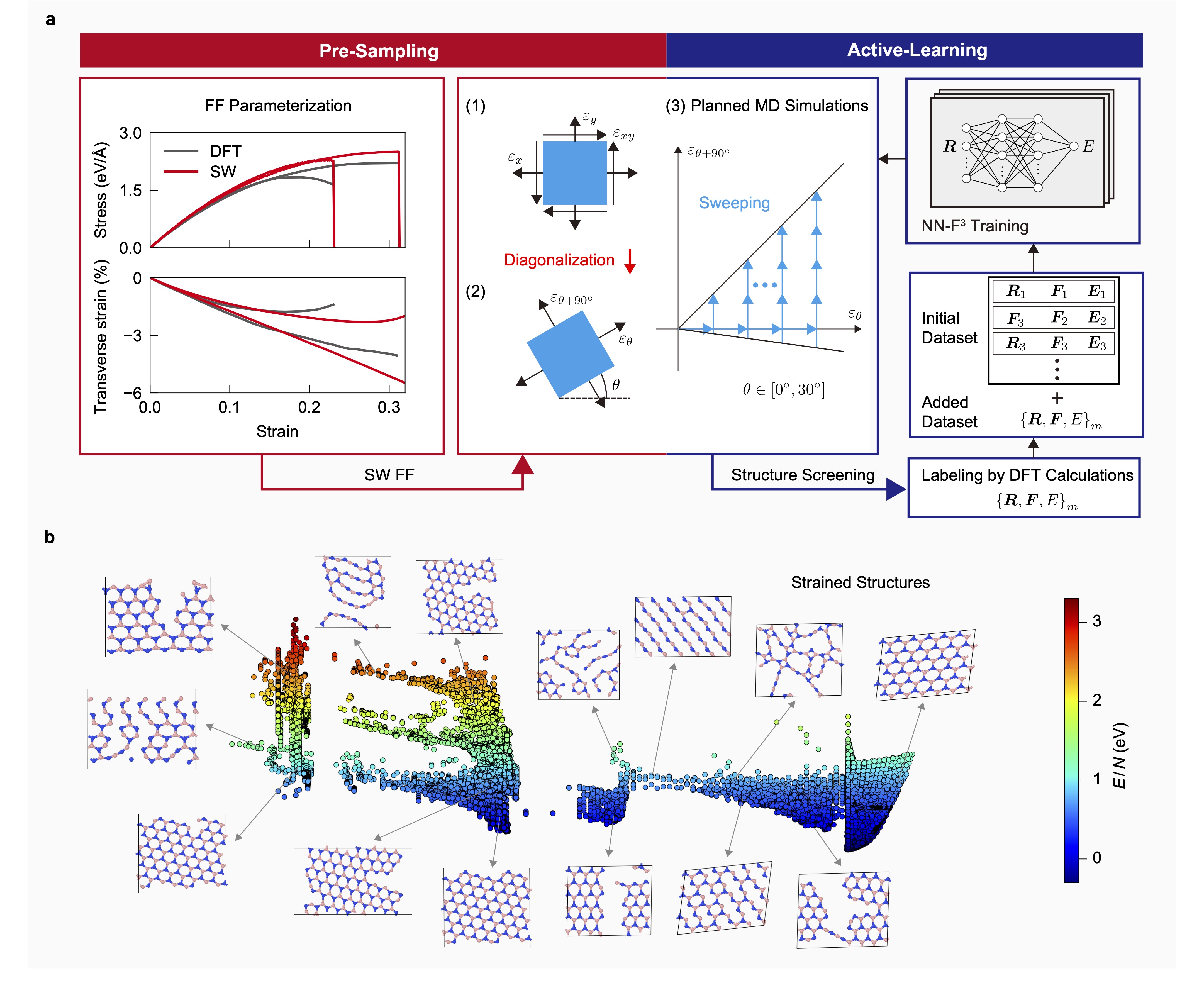}
\caption{
    The workflow of NN-F$^3$ development.
    \textbf{a}, In the pre-sampling step, stress-strain relations and transverse strains under uniaxial tension are calculated using DFT and SW FF. 
    An exploration process is then introduced in the active-learning step.
    The principal strain components are calculated.
    MD simulations are then performed by sweeping the parameters $\theta$, $\varepsilon_\theta$, and $\varepsilon_{\theta+90^\circ}$.
    Data are labeled by DFT calculations and then added to the dataset for NN-F$^3$ training in the active-learning step.
    \textbf{b}, Sketch map of the complete dataset ($97,271$ structures) used to train NN-F$^3$ for h-BN. 
    The color represents the total energy of the corresponding structures.
}
\label{Fig_2}
\end{figure*}

\subsection*{Pre-sampling of the strain states (\emph{Planning})}
The training dataset consists of atomic-level structures labeled by total energies, atomic forces, and virial coefficients obtained from DFT calculations.
A high-quality initial dataset can expedite the convergence and significantly reduce time spent on the subsequent active-learning process.
Particularly in our NN-F$^3$, we need to traverse the 3D space of strain states to derive the stress-strain relations.
For efficient sampling in the initial structures of the 2D crystals, the Stillinger-Weber (SW) FF\,\cite{hossain_2018} is parameterized based on DFT calculation results 
and used in the following MD simulations to produce the atomic-level structures in the construction of the initial dataset.
The SW FF $V=V_2+V_3$ includes both two-body and three-body parts, which are
\begin{align}
    V_2\left(r_{ij}\right)=&A\left(B\frac{\sigma^4}{r_{ij}^4}-1\right)\exp{\left(\frac{\sigma}{r_{ij}-r_{\rm c}}\right)}\\
    V_3\left(\theta_{ijk}\right)=&\lambda\left(\cos\left(\theta_{ijk}-\theta_0\right)\right)^2\exp\left(\frac{2\gamma\sigma}{r_{ij}-r_{\rm c}}\right),
\end{align}
respectively\,\cite{hossain_2018}. The equilibrium lattice constants and stress-strain relations of graphene and h-BN under uniaxial (along both the zigzag and armchair directions) and biaxial tension are included in parameterizing SW.
Specifically, the two-body parameter $B$ in SW is determined by the equilibrium lattice constants \cite{hossain_2018}.
Other two-body parameters are then fitted to the energy-strain relations obtained under the biaxial tension test that does not include the contribution of the three-body term by assuming $\theta_0=120^\circ$.
The honeycomb lattice is skewed under uniaxial tension tests, which define the three-body parameters.
The range of strain explored using SW is set to be broader than that in the referenced DFT calculations for efficient sampling (Fig.\,\ref{Fig_2}a).

Planned MD simulations are then carried out to sample the large strain configurations (Fig.\,\ref{Fig_2}a).
The basal-plane principal strain values ($\varepsilon_\theta$, $\varepsilon_{\theta+90^\circ}$) and the orientation angle ($\theta$), which can be obtained by diagonalizing the strain tensor, are controlled to sweep the space of strain states.
The values of $\theta$ are restricted to the range $\left[0^\circ,30^\circ\right]$ following the lattice symmetry. 
We traverse the 3D space of parameters ($\theta$, $\varepsilon_\theta$, $\varepsilon_{\theta+90^\circ}$) in the MD simulations and uniformly sample the structures.

The structures generated by SW simulations are then fed into DFT calculations, yielding a set of $12,022$ data frames of atomic positions, total energies, atomic forces, and virial coefficients (see Methods for details).
The predictions of NN-F$^3$ trained using this initial dataset are shown in Supplementary Fig.\,4. 
The stress-strain relations agree well with the DFT calculations, and the mean absolute error (MAE) is as low as $21.07$ meV/\AA$^2$.
However, stress prediction approaching the peak strain and the Poisson ratio at large strain deviate from the reference DFT calculations because under large strain, which is attributed to the limited sampling by structures generated simulations using SW FF.

\subsection*{Active learning for rare events (\emph{Exploration})}
To account for the highly-distorted structures at the crack tips and the undercoorditation nature of cleaved edges, the initial dataset needs to be expanded for improved predictions of the atomic forces.
We adopt an active-learning strategy (\emph{Training-Exploration-Labeling}\,\cite{dpgen}) to explore the most relevant structures iteratively based on a predefined criterion of the deviation in atomic forces (Fig.\,\ref{Fig_2}a).

Firstly, $4$ NN-F$^3$s are trained based on the current dataset but using different seeds for random-number generation in initializing the NN parameters.  
$1$ of the $4$ NN-F$^3$ is used to run MD simulations using the Atomic Simulation Environment (ASE)\,\cite{ase} to generate trajectories and compute the atomic forces.
The loading conditions are the same as those in the pre-sampling process.
The \emph{Query by Committee}\,\cite{smith2018less} algorithm is then used to screen the structures.
We select structures of graphene or h-BN monolayers with a maximum standard deviation (SD) of atomic forces (over the $4$ NN-F$^3$s) exceeding $0.1~\mathrm{eV/\AA}$ for the subsequent DFT calculations.
Atomic-level structures containing crack tips and open edges are identified by atoms with coordination numbers less than $3$.
These structures are selected with a maximum SD exceeding $0.2$ eV/\AA.
The screened structures are labeled by DFT calculations and the results are added to the product dataset. 
As a natural outcome of the MD exploration process, structures containing crack tips emerge and are included in the dataset.

\begin{figure*}[t]%
\centering
\includegraphics[width=\textwidth]{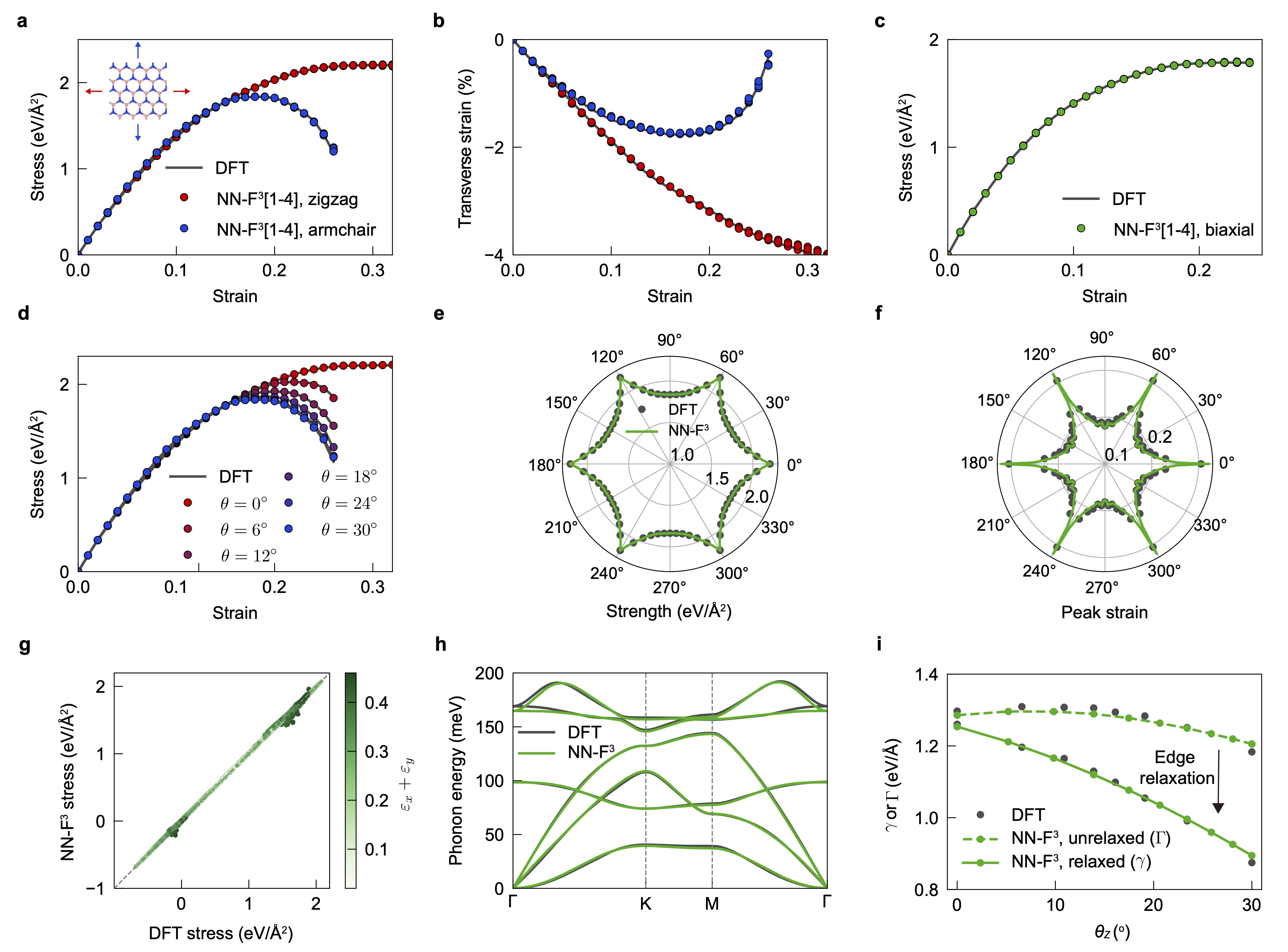}
\caption{
Performance of NN-F$^3$ for h-BN by comparison with DFT predictions.
\textbf{a}-\textbf{c}, Stress-strain relations under uniaxial tension along the zigzag and armchair directions (\textbf{a}, \textbf{b}),  and under biaxial tension (\textbf{c}) for all the $4$ NN-F$^3$s. 
\textbf{d}-\textbf{f}, Stress-strain relations (\textbf{d}), strength (\textbf{e}) and peak strain (\textbf{f}) under uniaxial tension along different lattice orientations.
\textbf{g}, Performance of stress predictions with $\varepsilon_{x}$ and $\varepsilon_{y}$ in the range of $\left[0,0.24\right]$. 
\textbf{h}, Phonon spectrum. 
\textbf{i}, Energy densities of unrelaxed and relaxed edges along different lattice orientations ($\theta_\mathrm{Z}$, measured from the zigzag motif).
}\label{Fig_3}
\end{figure*}

\subsection*{Performance of NN-F$^3$}
Our methods are validated for both graphene and h-BN, which is illustrated here using h-BN as an example.
A total of $8$ active-learning iterations are performed to reach the convergence.
The maximum SDs of all structures are below the threshold of $0.1~\mathrm{eV/\AA}$.
The final dataset consists of $97,271$ data frames, $45,621$ of which are structures containing crack tips and open edges.
The product dataset is represented by a sketch map (Fig.\,\ref{Fig_2}b), where the points are associated with structures in the training dataset. 
The position of each point is determined by the total energy normalized by the number of atoms and the distribution of coordination numbers.
The simulation snapshots added to the sketch map show the diversity of structures in the dataset, which include lattices at different strain states, cleaved edges, structures containing crack tips, as well as chains and net structures characterized during the fracture process.

The performance of NN-F$^3$ is summarized through the predicted energies, forces, and stress-stain relations (Fig.\,\ref{Fig_3}, Supplementary Fig.\,5). 
The MAEs of the energy per atom, the interatomic forces, and the in-plane stress are below $3.06$ meV/atom (Supplementary Fig.\,5a), $50.8$ meV/\AA (Supplementary Fig.\,5b) and $1.76$ meV/\AA$^2$  (Fig.\,\ref{Fig_3}a-g), respectively.
Fig.\,\ref{Fig_3}a-c demonstrates the consistency among the $4$ NN-F$^3$s and with DFT calculations regarding the stress-strain relations in uniaxial tension tests (in both zigzag and armchair directions) and biaxial tension.
Furthermore, the stress-strain relations of uniaxial tension along various directions (Fig.\,\ref{Fig_3}d), strengths (Fig.\,\ref{Fig_3}e), and peak strain (Fig.\,\ref{Fig_3}f) for all directions exhibit excellent agreement with the reference results obtained from DFT calculations.
For the equilibrium properties, the MAE of the phonon spectrum relative to the DFT results is $0.213$ meV. 
The energy densities of unrelaxed or relaxed edges show excellent consistency with DFT calculations, which are often used in the estimation of fracture toughness for brittle materials\,\cite{lawn_2004} (Fig.\,\ref{Fig_3}i).

\section{Results and Discussion}\label{sec3}

The capability of NN-F$^{3}$ for extreme mechanical processes is demonstrated by choosing two representative problems on the prediction of fracture pattern. Firstly, we explore the heteroatomic nature of h-BN and its effect on the edge cleavage processes. Secondly, we integrate NN-F$^{3}$ with the interlayer interaction between two twisted neighboring graphene layers to study the fracture of van der Waals (vdW) structures.

\subsection*{Fracture of h-BN}
\begin{figure*}[t]%
\centering
\includegraphics[width=\textwidth]{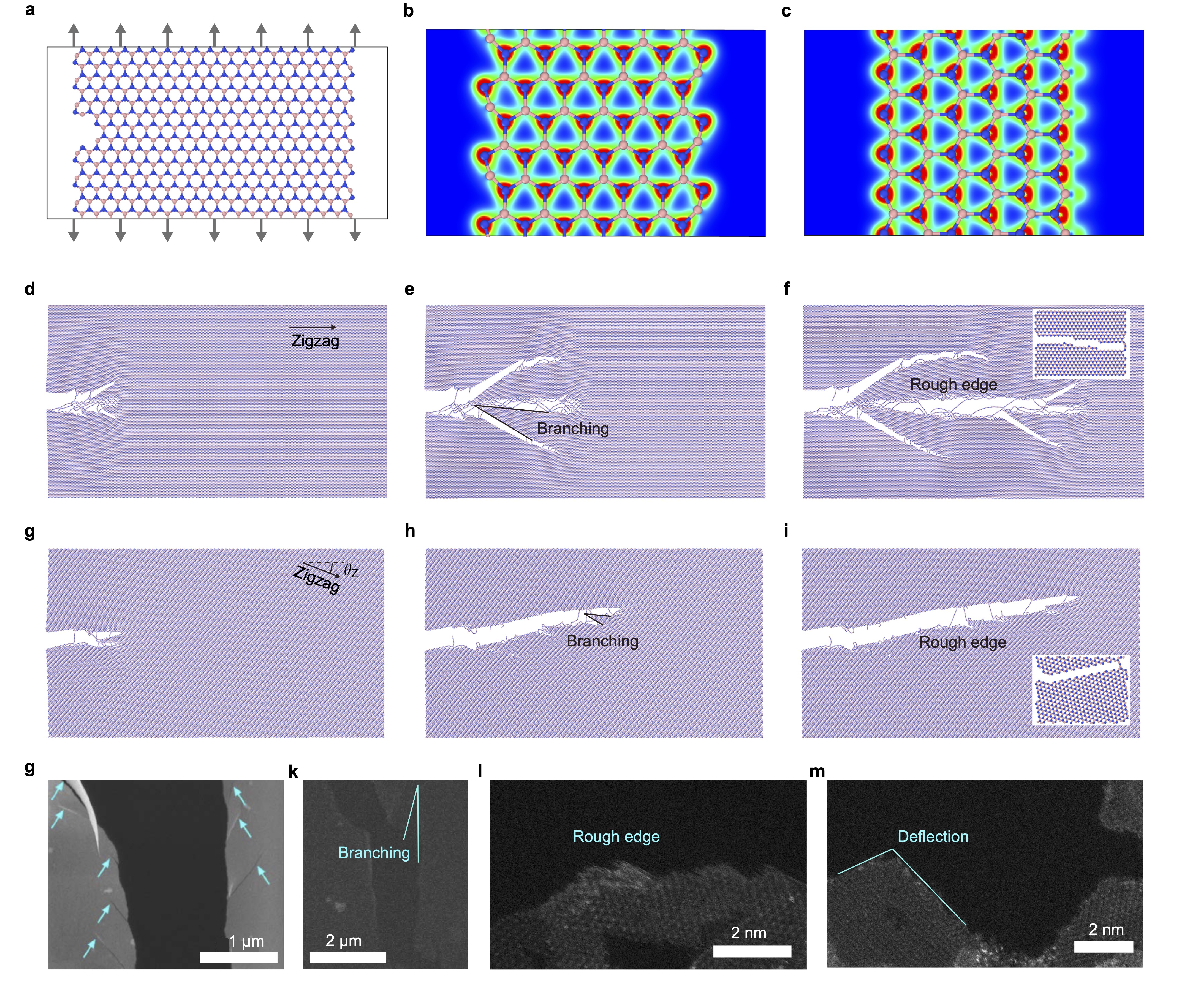}
\caption{
Fracture patterns of h-BN.
\textbf{a}, Geometry of the pre-crack and the loading condition of uniaxial tension.
\textbf{b}-\textbf{c}, Charge density distribution of armchair and zigzag edges of h-BN.
\textbf{d}-\textbf{i}, Fracture patterns under uniaxial tension along the zigzag direction ($\theta_\mathrm{Z} = 0^\circ$, \textbf{d}-\textbf{f}) and $\theta_\mathrm{Z} = 19.11^\circ$ (\textbf{g}-\textbf{i}). 
The insets in panels \textbf{f} and \textbf{i} show the fracture patterns of samples with reduced sizes (hundreds of atoms). 
\textbf{g}-\textbf{m}, Scanning electron microscopy (SEM) images of the fracture patterns, showing the features of crack branching, rough edges, and crack deflection~\cite{zhigong_2021}.
}\label{Fig_4}
\end{figure*}

Developing strong and tough materials has been a long-term goal in engineering sciences.
To resolve the conflict between material strength and toughness, intrinsic or extrinsic concepts such as crack deflection, microcracking, and fiber bridging are introduced\,\cite{ritchie_2011}.
2D crystals such as h-BN and graphene are known to feature superior strength and strain to failure~\cite{cao2020elastic,han2020large}.
Intrinsic toughening of h-BN was recently discovered and explained by the asymmetry of B and N edges cleaved during fracture (Fig.\,\ref{Fig_4}a), which is absent in graphene with a single composition of carbon element\,\cite{zhigong_2021}.
However, empirical FFs cannot capture the mixed nature of ionic and covalent bonding in h-BN (Fig.\,\ref{Fig_4}b,c), while first-principles studies of crack nucleation and growth are limited by the size of models.
Our NN-F$^3$ is thus used to explore the problem by offering simultaneously DFT-level accuracy and low computational costs (see Methods for details).

The fracture patterns of h-BN lattices are shown in Fig.\,\ref{Fig_4} d-f, g-i, where the uniaxial tensile load is applied in the direction with an angle of $\theta_\mathrm{Z} = 0^\circ$ and $19.11^\circ$ with the zigzag motif.
In contrast to the results of graphene, rough edges are identified from the MD simulation results, showing crack deflection and branching along the path of propagation at different length scales.
Similar characteristics are also observed in fracture patterns with $\theta_\mathrm{Z}=10.88^\circ$ or $30^\circ$ (Supplementary Fig.\,\,6). 
These features align well with the experimental evidence reported from \emph{in situ} SEM studies (Fig.\,\ref{Fig_4}g-m).
The instability of crack propagation was attributed to the asymmetry between the B and N sites at the edge of h-BN\,\cite{zhigong_2021}, which results in shear along the edge and local $K_{\rm II}$ components at the crack tip, deflecting the cracks\,\cite{cheng1990kii,zhigong_2021}. 
The $K_{\rm II}$ field is localized at the crack tip.
The calculated values of $K_\mathrm{II}/K_\mathrm{I}$ decreases significantly as the distance from the crack tip, $r$, increases (Supplementary Fig.\,\,7c). 
The asymmetry in $\mid\tau\mid$ and $\sqrt{r}\left[\tau\left(r, \theta\right)+\tau\left(r, -\theta\right)\right]$ confirm the localization of shear near the crack tip (Supplementary Fig.\,\,7).
Here $\tau$ is $x-y$ component of the stress tensor and $\theta$ are defined in the crack tip coordinate system in Supplementary Fig.\,\,7g.

It should be noted that the size effects in simulating fracture behaviors are crucial\,\cite{buehler_2010}.
The DFT calculations are usually limited to models with hundreds of atoms, while the NN-F$^{3}$ simulations here are carried out for more than $60$ thousands of atoms.
For comparison, MD simulations of a reduced-size model with $1,000$ atoms are carried out (inset of Fig.\,\ref{Fig_4}f, i and Supplementary Fig.\,6b.
The results fail to produce features such as the rough edges (inset of Fig.\,\ref{Fig_4}f) and crack deflection (inset of Fig.\,\ref{Fig_4}i) in the large-scale NN-F$^{3}$ simulations and highlight the significance of NN-F$^{3}$ in studying the problem of fracture.

Interestingly, atomic-level structures such as single-atom chains and net structures are observed in the simulations (Fig.\,\ref{Fig_4}d-f). 
The existence of BN chains was reported by experiments\,\cite{cretu2014experimental}, while the net structures have yet to be reported to the best of our knowledge.
To confirm the predictions, thermodynamic stabilities of the net structures are assessed by NN-F$^3$ and finite-temperature Born-Oppenheimer MD (BOMD) simulations.
Two single-atom BN chains spaced by $0.17$ nm are simulated at $T = 300$ K (Supplementary Videos\,1 and 2).
The simulation results consistently show attraction between the chains and the formation of net structures as a result. 
The computed formation energies of chain and net structures suggest that the net structures featuring an alternating arrangement of quaternion and octagon rings (Supplementary Fig.\,8e) possess lower energies ($0.885\,\mathrm{eV/\AA}$) compared to that of the BN chain ($0.920\,\mathrm{eV/\AA}$).
This result confirms the stability of the net structures. 

\subsection*{Fracture of twisted graphene bilayers}

\begin{figure*}[t]%
\centering
\includegraphics[width=\textwidth]{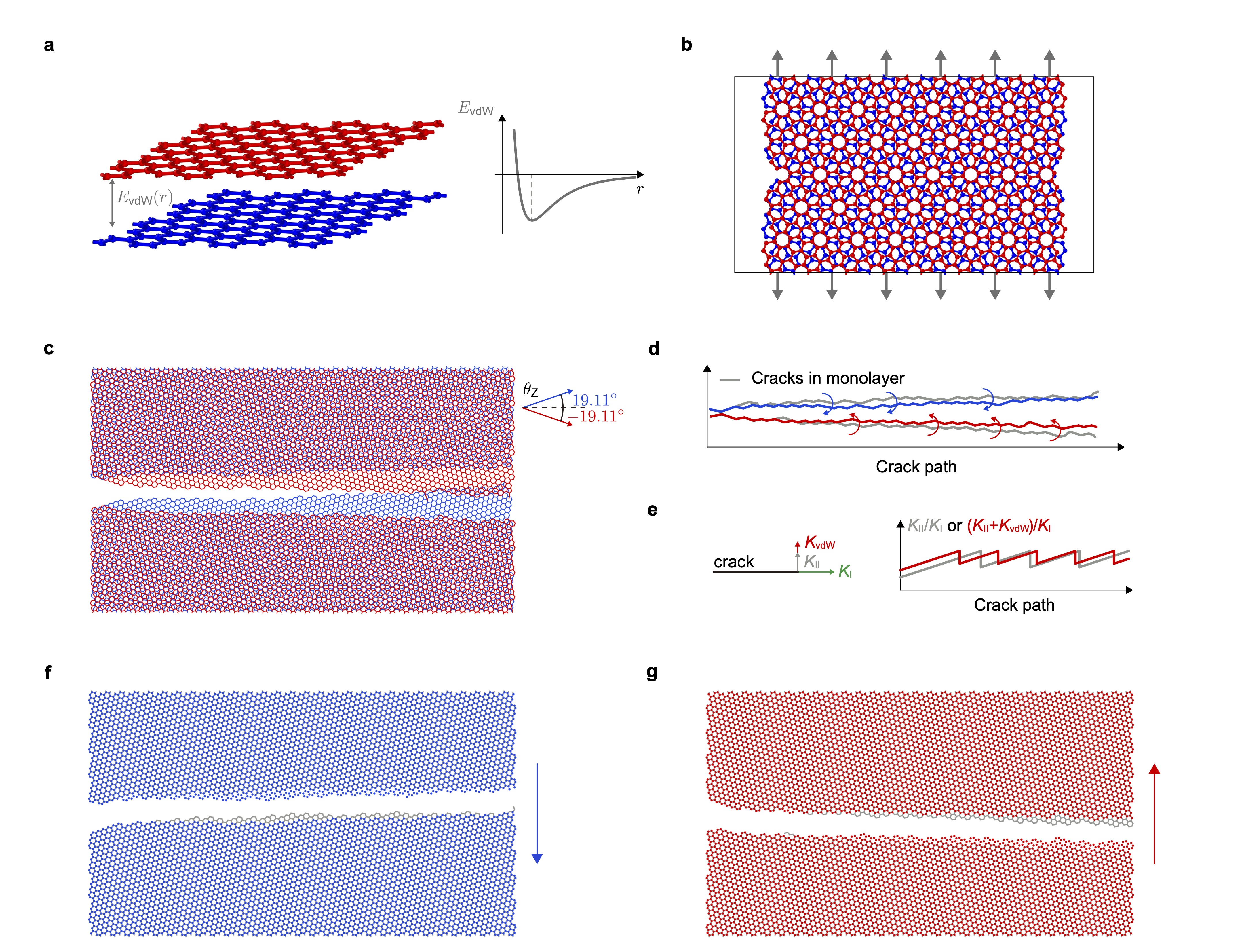}
\caption{
Fracture patterns of twisted bilayer graphene (TBG).
\textbf{a}, Structures of TBG and the Lennard-Jones (LJ) potential used to model the interlayer interaction.
\textbf{b}, Loading geometry and conditions.
\textbf{c}, Fracture patterns of TBG under uniaxial tension along lattice directions ($\theta_\mathrm{Z} = \pm19.11^\circ$ for the top and bottom layers, respectively).
\textbf{d}, The crack paths showing attraction between the cracks across the interface.
The crack paths of isolated single-layer graphene with the same lattice orientation are shown in gray lines.
\textbf{e}, Effective stress intensity factors of the inter-layer interaction ($K_{\rm LJ}$) that deflect the cracks.
\textbf{f}, \textbf{g}, Fracture patterns of the top (\textbf{f}) and bottom (\textbf{g}) graphene layers, respectively.
The fracture patterns of isolated single-layer graphene with the same lattice orientation are shown in gray color.
The arrows indicate the directions of attraction between the cracks due to interlayer interactions.
}\label{Fig_5}
\end{figure*}

NN-F$^{3}$ can be integrated with FFs for the interlayer interaction in multilayers or heterostructures of 2D materials.
Here we study the fracture of twisted bilayer graphene (TBG), which is made up of two layers of graphene stacked with a specific angle of rotation (Fig.\,\ref{Fig_5}a). 
The misalignment in lattice orientation between the two layers leads to the formation of moiré patterns (Fig.\,\ref{Fig_5}b).
Exotic quantum phases such as the correlated insulating phase, unconventional superconductivity, and the fractional Chern insulator phase were reported\,\cite{cao2018correlated,cao2018unconventional,yankowitz2019tuning,xie2021fractional}, making TBG a promising material for applications in electronics\,\cite{heikkila2022surprising}, optoelectronics\,\cite{wang2022polarization}, and quantum computing\,\cite{xie2021fractional}.
However, given the challenges associated with fabricating and implementing TBG-based devices, the failure or shaping, in a positive point of view, led by the fracture of TBGs remains a technical concern to be addressed\,\cite{huang2020large,mesple2021heterostrain}.
One of the distinct and interesting problems on this topic is the interaction between cracks in neighboring layers, which could potentially toughen the structures by shielding the stress field or modifying the crack paths\,\cite{lin2014step,arshad2023fracture}.
However, an accurate and efficient force field is necessary to predict the crack paths in reasonably sized models\,\cite{shi2023nonequilibrium}, to produce theoretical predictions that can be directly compared to experimental data\,\cite{feng_2022,qu_2022}.

In the absence of electrostatic interaction and polarization, only dispersion or vdW interaction needs to be considered for interaction between graphene layers.
In MLFFs, instead of increasing the cutoff of interatomic interaction in FF construction, a dispersion term (e.g., $r^{-6}$\,\cite{deringer2020general,wen2019hybrid}) is commonly added for the sake of convenience\,\cite{anstine2023machine}.
Alternatively, a short-range descriptor can be added to predict the effective atomic Hirshfield volume that is used for dispersion correction\,\cite{muhli2021machine}.
In this work, we combine NN-F$^3$ with the $12-6$ LJ potential\,\cite{stuart2000reactive} to model TBGs (Fig.\,\ref{Fig_5}a).  
The approach can be extended to, for example, the Kolmogorov-Crespi (KC) potential\,\cite{kolmogorov2005registry} to further include short-range Pauli repulsion between overlapping $\pi$ orbitals of adjacent layers.

Uniaxial tension is applied for the TBG along the zigzag directions ($\theta_\mathrm{Z} = \pm19.11^\circ$ for the top and bottom layer, respectively (Fig.\,\ref{Fig_5}c).
Simulations of a graphene monolayer with $\theta_\mathrm{Z}$ of $19.11^\circ$ are also performed for comparison.
The results show that the cracks in the bilayer graphene approach each other as a result of attraction between the cracks across the vdW interface (Fig.\,\ref{Fig_5}d, f, g).
The modified crack paths indicate toughening in the TBG in comparison with single graphene layers.

Additional fracture tests are performed for TBGs with different lattice orientations (Supplementary Fig. 9). 
The interaction between cracks in neighboring graphene layers is measured by the difference in the number of kinks along the crack paths between TBG and the single layers with the same lattice orientation (Supplementary Table 1).
The results show that the interaction is weakened as $\theta_\mathrm{Z}$ deviates from $\pm19.11^\circ$, and becomes negligible for large deviation or two aligned cracks (e.g., $\theta_\mathrm{Z}=\pm17.48^\circ$, Supplementary Fig. 10).
This phenomenon may be attributed to the competition between the stress field in the opening mode (mode \uppercase\expandafter{\romannumeral1}, quantified by the stress intensity factor or SIF, $\Kone$) that preserves the crack direction, the stress field in the shear mode (mode \uppercase\expandafter{\romannumeral2}, $\Ktwo$) that deflects the crack, and the weak interlayer interaction that adds an additional driving force, $K_{\rm vdW}$.
As the ratio of $\Ktwo/\Kone$ approaches a critical value at which crack deflection can be activated, the interlayer interaction significantly modifies the crack paths (e.g., $\theta_\mathrm{Z}=\pm19.11^\circ $).
Otherwise, the effect of interlayer interaction becomes negligible (e.g., $\theta_\mathrm{Z}=\pm27.46^\circ $) (Fig.\,\ref{Fig_5}e).

The interlayer vdW interaction was reported to be crucial for complex fracture behaviors of 2D materials\,\cite{ni2022fracture}.
For example, cracks propagate along dissimilar paths in trilayer graphene as a result of the interlayer slippage\,\cite{jang2017asynchronous}.
In the time domain, asynchronous fracture in bilayer graphene is also observed\,\cite{lin2014step}.
Crack paths in bilayer $\mathrm{MoS_2}$ are closely related to the interlayer stacking order and in-plane loading conditions\,\cite{jung2018interlocking}. 
The above phenomena are identified in our simulations of TBG.
Interlayer interactions such as H-bonding, electrostatic, and covalent bonding via functionalization can enhance the load transfer between the 2D layers and modify the crack propagation behaviors, resulting in strengthening and toughening effects \,\cite{cao2018nonlinear}.
The NN-F$^{3}$ reported here lays the ground for direct simulations of these behaviors.

\section{Conclusion}\label{sec4}
In this work, we combine pre-sampling and active learning to develop a neural network-based force field for fracture (NN-F$^3$). 
The framework takes into account large-strain effects (e.g., nonlinearity, anisotropy), bond breakage and (re)formation at the crack tips, and relaxation of the cleaved edges, all of which are closely related to fracture.
The high-fidelity NN-F$^3$ offers unprecedented DFT-level accuracy in exploring the multiscale nature of fracture as an example of the mechanical behaviors of materials under extreme conditions, which have not been achieved with other force fields in the literature.
The capability of NN-F$^3$ is demonstrated by modeling crack deflection and branching in h-BN and the cross-plane interaction between cracks in twisted bilayer graphene.
The results elucidated the underlying mechanisms of these processes that cannot be adequately predicted by the state-of-the-art models of interatomic interactions.
The use of NN-F$^{3}$ in simulating material failure can further improve the understanding and prediction by integration with the recently proposed deep learning models\,\cite{lew2021deep}.

\section*{Acknowledgments}\label{acknowledgement}
This study was supported by the National Natural Science Foundation of China through grants 11825203, 11832010, 11921002, and 52090032.
The computation was performed on the Explorer 100 cluster system of the Tsinghua National Laboratory for Information Science and Technology.

\section*{Data Availability}\label{data}
The data that support the findings of this study are available upon reasonable request from the authors.

\appendix
\section{DFT calculations}
Spin-polarized DFT calculations are performed using the Spanish Initiative for Electronic Simulations with Thousands of Atoms (SIESTA) package\,\cite{siesta} using numerical atomic orbitals (NAOs) at the double-$\zeta$-plus-polarization (DZP) level. 
Perdew-Burke-Ernzerhof (PBE) parameterization of the generalized gradient approximation (GGA) is used for the exchange-correlation functional\,\cite{PBE}.
Troulliere-Martins-type norm-conserving pseudopotentials are chosen for the ion-electron interactions\,\cite{troullier1991}. 
The cut-off energy for electron wave functions is $500$ Ry.
The $\mathbf{k}$-space is sampled by a $4\times4\times1$ Monkhorst-Pack grid for the $72$-atom model.
For the structures with open edges, sampling at the same $\mathbf{k}$-point density is used.
\section{Development of NN-F$^{3}$}
The training process of NN-F$^{3}$ adopts an active-learning (\emph{Training-Exploration-Labeling}) workflow.\\
DeePMD\,\cite{deepmd} is used to train the force fields.
The sizes of the embedding and fitting nets are $(25, 50, 100)$ and $(240, 240, 240)$.
The cut-off radius and the smoothing parameter are $6.0$~\AA~and \emph{rcut\_smth} $=5.0$~\AA, respectively. 
The batch size is $1$.
Adaptive moment estimation (Adam) optimization is performed for $3\times10^7$ steps to train the DP models.
The hyper-parameters \emph{start\_pref\_e}, \emph{start\_pref\_f}, \emph{limit\_pref\_e} and \emph{limit\_pref\_f} that control the weights of energy and force losses in the total loss function are set to $1.0$, $10.0$, $1.0$ and $10.0$, respectively.
The starting learning rate is $0.001$, which exponentially decays to $1.0\times10^{-8}$ at the end of the training.
$95\%$ of our dataset is used to train the model, and the rest is used for validation.
Atomic Simulation Environment (ASE)\,\cite{ase} is employed to conduct molecular dynamics (MD) simulations in the exploration process.
The rate of the strain sweeping processes is $1\times10^{-5}$ ps$^{-1}$.
In MD simulations, the structures are screened at each timestep.
DFT calculations using SIESTA are used to label the screened structures.
\section{MD simulations}
To simulate the fracture of h-BN, we use the Large-scale Atomic/Molecular Massively Parallel Simulator (LAMMPS) \cite{lammps}.
The size of h-BN samples is approximately $60~\mbox{nm}\times30~\mbox{nm}$ and the total number of atoms ranges from $65,519$ to $66,975$.
The Nos\'e-Hoover thermostat is used for temperature control, and the damping constant is $0.1$ ps.  
The strain rate used in tensile tests is $1\times10^{-3}$ ps$^{-1}$.
To simulate twisted bilayer graphene (TBG), we combine NN-F$^3$ with the Lennard-Jones (LJ) potential in ASE.
The size of TBG samples is approximately $22.5~\mbox{nm}\times10~\mbox{nm}$, and the total number of atoms ranges from $14,212$ to $15,379$.

\bibliography{bibliography.bib}

\end{document}